# BROADCAST SEARCH IN INNOVATION CONTESTS: CASE FOR HYBRID MODELS

Thomas Gegenhuber
Johannes Kepler University (JKU)
Austria
thomas.gegenhuber@jku.at

Marko Hrelja
Ryerson University
Canada
mhrelja@ryerson.ca

**ABSTRACT**

Organizations use broadcast search to identify new avenues of innovation. Research on innovation contests provides insights on why excellent ideas are created in a broadcast search. However, there is little research on how excellent ideas are selected. Drawing from the brainstorming literature we find that the selection of excellent ideas needs further investigation. We propose that a hybrid model may lead to selection of better ideas. The hybrid model is a broadcast search approach that exploits the strengths of different actors and procedures in idea generation and the selection phase.

## I. INTRODUCTION

Innovation is important to gain a competitive advantage and benefits from multiple perspectives. Organizations may seek those multiple perspectives outside of the organization by using broadcast search. InnoCentive is a typical example for broadcast search: An organization attempts to solve a problem by posting a challenge for a group of different, a priori unidentified agents to generate solutions. These agents do not have access to the contributions of their peers. The organization selects the "best" idea and awards the winnner(s). The broadcast search model can also be applied to tap into multiple perspectives within an organization as the example InnoCentive@work shows. The literature provides answers to the question, "how does broadcast search lead to the generation of excellent ideas?". Still, the question, "how excellent ideas are selected", has not attracted as much attention. We will conceptually overlap findings of the brainstorming literature with broadcast search. We analyze different types of nominal groups and propose that a hybrid model that exploits the strengths of different actors and procedures works best in idea generation and selection. To fully capture the hybrid model, we further attempt to extend the "Collective Intelligence Genome Framework" of Malone et. al (2009; 2010).

This conceptual paper is structured as follows: First, we will briefly review "Open Innovation", the "Collective Intelligence Genome Framework" and the brainstorming literature. Next, we will establish the link between crowdsourcing applications such as contests and nominal groups. We selected three different group structures of the brainstorming literature and discuss their performance in idea generation and selection. Finally, drawing upon our findings, we will theorize about the hybrid model and discuss its implications.

## II. LITERATURE REVIEW

### IIa. Contests in Open Innovation and Crowdsourcing

Chesbrough (2003) put forward the idea of "Open Innovation". This paradigm highlights the importance to bring external knowledge into an organization but also identifies new strategies to exploit the organization's existing intellectual property. Thus, knowledge can flow inbound from outside into the organization (i.e. license in, spin in, acquire) or outbound, from inside to the outside of the organization, in the form of licensing, spin outs or divesting (Chesbrough, 2003; 2006). Schenk and Guittard (2011) state that from the perspective of "Open Innovation", a crowdsourcing application such as a contest (Howe, 2008), would be considered as another source to create an inbound knowledge flow to the organization.

Lakhani et. al (2007) & Jeppesen and Lakhani (2010) define broadcast search as a search process that opens problem information to self-selected outsiders. In an empirical study on InnoCentive, a typical example for broadcast search in the form of a winner-take-all contest, Lakhani et. al (2007) found "that the broadcast of problem information to outside scientists results in a 29.5% resolution rate for scientific problems that had previously remained unsolved inside the R&D laboratories of well-known science-driven firms" (2007: 4).

| Example | What | | Who | Why | How |
|---|---|---|---|---|---|
| InnoCentive | Create | Scientific Solutions | Crowd | Money | Contest |
| | Decide | Who gets reward | Management | Money | Hierarchy |

*Table 1: InnoCentive Genome (Malone et. al, 2010)*

But what is responsible for the creation of excellent solutions? Jeppesen and Lakhani (2010), who locate their work in the field of distributed/open innovation, report that marginality is an asset for problem solving in broadcast search. Marginality means that solvers have the advantage of not being burdened with assumptions and bring in new perspectives and heuristics to solve the problem. Jeppesen and Lakhani (2010) identify two types of marginality: technical marginality (coming from a different field than the problem) and social marginality (distant from the own professional community). Villarroel and Reis (2010) add in their empirical study about an internal innovation prediction market that rank marginality (lower position of an employee) and site marginality (great distance to the headquarters) are related with better innovation performance. We can summarize that broadcast search leads to generation of excellent ideas. Yet, the question that remains is, how are excellent ideas/solutions selected?

## II b. Brainstorming Literature

Kavadias and Sommer (2009) consider brainstorming and nominal groups as "multiagent searches for a solution to a problem" (Kavadias and Sommer, 2009: 1899). Kavidias and Sommer (2009) refer in their work to the innovation contests of Terwisch and Xu (2008). The similarities between nominal groups and innovation contests: In nominal groups, the members create their ideas individually and do not have access to the ideas of other group members. Principally, the broadcast search approach of innovation follows the same logic. Many scholars drew upon the brainstorming literature to discuss web-based idea-generation and elicitation systems (c.f. Dalal et. al, 2011; Krieger and Wang, 2008; Muhdi et. al, 2011), which we see as a support for our attempt to employ an analogy between nominal groups and broadcast search such as a contest.

## II c. Collective Intelligence Genome Framework

Malone et al. (2009) broadly define collective intelligence as "groups of individuals doing things collectively that seem intelligent" (2009: 2).

Malone et al. (2010) put forward the "Collective Intelligence Genome Framework" that captures the who, what, how and why of a crowdsourcing platform. Table 1 provides an example of how Malone et al. (2010) applied their framework to InnoCentive. We will apply the "Collective Intelligence Framework" in Section V to capture the process of idea generation and selection of a crowdsourcing application.

## III. PROCESS STRUCTURE OF CONTESTS AND NOMINAL GROUPS

First, we will establish the link between crowdsourcing efforts, such as contests, and nominal groups. Geiger et. al (2011) propose a taxonomy to classify crowdsourcing initiatives. The authors identify four characteristics of crowdsourcing processes: aggregation of contributions (integrative or selective), accessibility of peer contributions (modify, assess, view, none), remuneration for contributions (fixed, success-based or none), and preselection of contributors (qualification-based, context-specific, both, or none). Innovation contests like InnoCentive fall into the category of broadcast search (Howe, 2008; Jeppesen & Lakhani, 2010; Brabham, 2011). Following the taxonomy of Geiger et. al (2011), selective crowdsourcing without crowd assessment and selective sourcing with crowd assessment represent the broadcast search category. Table 2, an extract taken from Geiger et. al (2011:11) shows practical applications of broadcast search and clusters them based on the four process characteristics.

In selective crowdsourcing, crowd members individually create ideas. The organization can select the most favorable idea from the created set of options (Schenk and Guittard, 2010). Selective crowdsourcing is analogous to nominal groups. In contrast to interactive groups, group members in nominal groups create the ideas individually. After the idea generation phase, the ideas are pooled either by selection through individuals (Rietzschel et. al, 2006) or through assessment and selection (voting or rating) of an interactive group (Delbecq et. al., 1975).

| Cluster of process types | Crowdsourcing examples with the same process characteristics | Aggregation of contributions | Accessibility of contributions | Remuneration for contributions | Preselection of contributors |
|---|---|---|---|---|---|
| Selective sourcing without crowd assessment | Netflix Prize, InnoCentive Challenge Center, 99designs (private contests), Brainrack, Calling All Innovators, Crowdspring (private contests), Designenlassen.de (private contests), idea bounty | Selective | None | Success-based | No |
| | 99designs (public contests), Crowdspring (public contests), Designenlassen.de (public contests) | Selective | View | Success-based | No |
| Selective sourcing with crowd assessment | Atizo (Atizo Community), Cisco I-Prize, Threadless | Selective | Assess | Success-based | No |
| | Atizo (Own Community), InnoCentive@Work | Selective | Assess | Success-based | Context-specific |
| | Dell IdeaStorm | Selective | Assess | No | No |

*Table 2: Selective crowdsourcing process types of Geiger et. al (2010)*

The latter concept is the nominal group technique (NGT) (Delbecq et. al, 1975). The NGT process is as follows: The members of the NGT session sit together in one room and first write up individually their ideas on a sheet of paper. In the next step, everyone presents his/her own idea and the group discusses each idea. Finally, all members vote on the ideas by applying ranking or rating methods (Delbecq et. al, 1975).

In another approach, group members create the ideas individually; after the individual idea generation phase, the group works together to assess and modify the ideas. This combination of individual and group processes is called a 'hybrid' (Girotra et. al, 2010). Girotra et. al (2010) refer to prior literature (i.e. Robbins and Judge, 2006) which reports that a hybrid approach combines the merits of individual nominal and interactive brainstorming group processes.

*Table 3: Group Structures*

Table 3 summarizes the features of the different group structures by using the process types of Geiger et. al (2011).

## IV. ANALYSIS OF GROUP STRUCTURES

We have conceptually overlapped crowdsourcing applications, such as contests and nominal group structures. In this section we will discuss the performance of each group structure in the creation and selection of ideas.

## IV A. Nominal Groups

Kavadias and Sommer (2009) draw upon psychology literature to summarize that interactive groups suffer from

- evaluation apprehension,
- production blocking, and
- free riding.

|  | **Nominal Group (Rietzschel et. al, 2006)** | **NGT (Delbecq et. al, 1975)** | **Hybrid (Girotra et. al, 2010)** |
|---|---|---|---|
| **Aggregation of contributions** | Selective | Selective | Selective |
| **Accessibility of contributions** | None | Assess | First None, then Assess and Modify |
| **Remuneration** | Fixed | Fixed | Fixed |
| **Pre-selection of contributors** | Context-specific | Context-specific | Context-specific/qualification-based |

Rietzschel et. al (2006) compared in their experimental study the productivity (number of ideas, originality, feasibility) of nominal groups with interactive groups. Nominal groups in the study of Rietzschel et. al (2006) created and selected the ideas individually. The study found that nominal groups created more ideas than interactive groups and the ideas of nominal groups were more original. In contrary, the ideas of interactive groups were more feasible. Nevertheless, both group structures lack the capability of selecting the best ideas. Rietzschel et. al (2006) refer to Simonton (2003) who reports that individuals are not good in assessing the potential of their own ideas and that this inability does not change over the course of their academic careers. Rietzschel et. al (2006) assume that a 'hybrid' might be more successful for the tasks of idea generation and selection: "It is possible that a combination of nominal and interactive idea generation and selection would yield optimal results on both tasks" (Rietzschel et. al, 2006: 250-251).

### IV B. Nominal Group Technique (NGT)

Delbecq et. al (1975) describe NGT as an instrument to gather heterogeneous group members to pool their judgments for creative decision-making. NGT groups have a proactive search behavior and balance social and task oriented roles (Delbecq et. al, 1975). NGT divides the process into two phases; the idea generation phase and an evaluation phase. Mair and Hoffmann (1964) "suggest that one type of group process should be used to generate information and another type to reach a solution" to reduce ambiguity of the group "about differences in decision-making phases" (Mair and Hoffmann, 1964 in Delbecq et. al, 1975: 9). In the evaluation phase, the NGT method puts an emphasis on individual voting trough rank ordering or rating. Van den Ven (1974) found that NGT groups created twice as many ideas than interactive groups. Although Delbecq et. al (1975) suggest, that voting mechanisms lead selection of better ideas, the empirical support for the NGT is not "uniformly favorable" (Bartunek and Murninghan, 1984).

### IV C. A Hybrid Model

In their experimental study, Girotra et. al (2010), used a hybrid model. As mentioned previously, the participants created the ideas in a nominal group setting, the assessment and selection of the ideas took place in an interactive manner. Girotra et. al (2010) proposed following benchmarks for the identification of the best ideas: the average quality of ideas generated, the number of ideas generated, the variance in the quality of ideas generated and the ability to select the best idea. To measure the performance Girotra et. al (2010) assembled a team of students that were not involved in the idea generation process. This group rated all ideas (scale from one to ten - the highest value) by using a web-based tool to measure the business value. For the purchase intent, Girotra et. al (2010) invited potential customers to participate in a web-based survey. Gathering a diverse group for performance measurement seems to be a more accurate method than Rietzschel et. al (2006, 2010), who measured the performance by using a trained rater. Girotra et. al (2010) conclude that hybrid groups create more and better ideas. Girotra et. al (2010) found that the performance of the hybrid group was better in discerning the quality of ideas but this did not lead to the selection of better ideas.

### V. DISCUSSION

The discussion of the three different group structures in Section IV indicate that a hybrid model may lead to the creation of the best ideas, although the findings suggest that even a hybrid lacks the capability to select the best ideas. To succeed in both tasks, we think it is necessary to reconceptualize the hybrid. Thus, we interpret the hybrid model as an attempt to best utilize the strengths of different actors and procedures in idea generation and the selection phase in a broadcast search (c.f. Girotra et. al, 2010; Rietzschel, 2006). For idea generation, having a nominal group process helps avoid some of the pitfalls of idea generation (i.e. groupthink). Interactive groups may be better to tackle a cross-functional problem that has a medium complexity (Kavadias and Sommer, 2009). A good example for a hybrid is Deloitte's "Innovation Quest" (c.f Terwiesch and Ulrich, 2009: 20), which uses a combination of nominal and interactive processes in a multi-round contest.

*Proposition 1:* A hybrid model leads to the creation and selection of better ideas.

Rietzschel et. al (2010) conclude that selection performance improves if participants are instructed to select most creative ideas. It is important to note that the finding of Rietzschel et. al (2010) is based on an experiment where the participants had to select ideas, which they did not generate themselves. Girotra et. al (2010) suggest that "irrespective of group structure, the ability of idea generators to evaluate their own ideas is extremely limited, and is perhaps compromised by their involvement in the idea generation step" (2010: 600).
Thus we propose that agents, who create the ideas, should be separated from the selection phase.

Separation reduces the involvement-effect and should result in a better selection of ideas. Many instruments can increase the separation. A group of experts with domain knowledge can judge ideas or the crowd can assist in the selection process (c.f Girotra et. al, 2010). Again, we will use Deloitte's Innovation Quest to provide an example. After individuals created the ideas individually, a group of domain experts of the organization selects ideas for the next round. In the second round, the employees might team up with others to win the challenge. In the final phase, all employees are invited to vote and comment on the ideas, which is one criterion for selecting the winners (Terwisch and Ulrich, 2009).

*Proposition 2: Separation from idea generation leads to selection of better ideas.*

The "Collective Intelligence Genome Framework" of Malone et. al (2010) helps us to understand the mechanisms of a hybrid. The beginning of this paper showed how Malone et. al (2010) applied their framework to InnoCentive. InnoCentive has a nominal setting - solvers create their ideas individually and cannot access the production of other solvers. The organization that seeks a solution assigns a group of experts with domain knowledge, presumably the researchers of the R&D lab, to select the best idea. In contrast, "Deloitte's Innovation Quest" allows the assessment of peer contributions. Applying the "Collective Intelligence Genome Framework," both examples would have a contest gene. If the contest gene occurs in combination, (e.g. with voting), one can assume that it is a hybrid.
However, drawing upon the impact of nominal and interactive group structure on collective intelligence processes, we think it may be useful to make the distinction between public and private contests.

It may be beneficial for analyzing hybrids to make the accessibility of the crowd to the total production of a crowdsourcing effort more clear by adding a public and private extension to the contest gene.

Table 4 provides a definition for each gene and an appropriate example:

| Gene | Definition | Example |
| --- | --- | --- |
| **Contest-Public** | The crowd has access to the total production of each crowd member | Deloitte Innovation Quest, 99Designs (public contests) |
| **Contest-Private** | The crowd has no access to the total production of each crowd member | InnoCentive, 99Designs (private contest) |

*Table 4: Adaption of the contest gene*

*Proposition 3: Adaption of the "Collective Intelligence Genome Framework" to include a public and private contest gene.*

We apply these new genes to the previous example of the "Deloitte's Innovation Quest" in Table 5. Note that we also include the genes added by Wise et. al (2010).

## VI. CONCLUSION

This conceptual paper attempts to shed light on the question, "how are best ideas selected in broadcast search?". We propose that a hybrid model may lead to the selection of better ideas. The hybrid model is a broadcast search model that exploits the strengths of different actors and procedures in idea generation and the selection phase.

*Table 5: New contest gene applied to Deloitte's Innovation Quest*

| What | | Who | Why | How |
| --- | --- | --- | --- | --- |
| Create | Submit ideas electronically | Crowd | Money, Love, Glory Interest | Contest-Private |
| Decide | Select ideas for the next round | Management (Domain Experts) | Money, Love, Interest | Hierarchy |
| Create | Build groups and improve ideas | Crowd | Money, Love, Glory Interest | Contest-Public |
| Evaluate | Vote on the best ideas | Crowd | Love, Interest, Money | Voting |
| Decide | Determine winners | Mgmt. | Money, Love, Glory, Interest | Hierarchy |

By extending the "Collective Intelligence Framework" of Malone et. al (2010) we hope to make the identification of hybrids in innovation contests more clear for future research. However, there are many open questions. What is the right combination in the hybrid (e.g. creation in a nominal group setting, selection trough experts and prediction markets)? How does a broadcast search within the organization or a broadcast search outside the organization influence the design of a hybrid? What is the impact on multi-round-contest on the creation and selection of ideas? In the future, we hope to dig into the dynamics of hybrids and how the innovative output of collective intelligence applications is maximized.